\documentclass[11pt]{article}
\usepackage{fullpage}
\usepackage{bookman}

\usepackage{multirow}
\usepackage[normalem]{ulem}
\useunder{\uline}{\ul}{}

\usepackage{forest}
\usepackage{import}
\usepackage{xspace}
\usepackage{dirtree}
\usepackage{hyperref}
\usepackage[most]{tcolorbox}

\usepackage{enumitem}

\usepackage{booktabs}

\usepackage{cleveref}
\crefformat{section}{\S#2#1#3}
\crefname{figure}{Figure}{Figures}
\crefname{table}{Table}{Tables}
\crefname{listing}{Listing}{Listings}
\crefname{theorem}{Theorem}{Theorems}
\crefname{thm}{Theorem}{Theorems}
\crefname{lemma}{Lemma}{Lemmata}
\crefname{equation}{Eqt.}{Eqts.}
\crefname{appendix}{Appendix}{Appendices}
\crefformat{Grammar}{Grammar #1}

\newcommand{\code}[1]{\texttt{{\small #1}}}

\newcommand{\PU}{Purdue University\xspace}
\newcommand{\LUC}{Loyola University Chicago\xspace}
\newcommand{\TFMG}{TensorFlow Model Garden\xspace}
\newcommand{\TF}{TensorFlow\xspace}
\newcommand{\MG}{Model Garden\xspace}
\newcommand{\SIG}{Special Interest Group\xspace}
\newcommand{\TFDS}{TensorFlow datasets\xspace}

\newcommand{\CPU}{Central Processing Unit\xspace}
\newcommand{\GPU}{Graphics Processing Unit\xspace}
\newcommand{\TPU}{Tensor Processing Unit\xspace}
\newcommand{\ML}{machine learning\xspace}
\newcommand{\DL}{deep learning\xspace}

\newcommand{\rep}{reproducibility\xspace}
\newcommand{\NLP}{natural language processing\xspace}
\newcommand{\CV}{computer vision\xspace}
\newcommand{\OD}{object detection\xspace}

\newcommand{\VM}{virtual machine\xspace}

\newcommand{\ie}{\textit{i.e.,}\xspace}
\newcommand{\eg}{\textit{e.g.,}\xspace}
\newcommand{\etal}{\textit{et al.}\xspace}

\newif\ifDEBUG
\DEBUGtrue

\ifDEBUG
    \newcommand{\vishnunote}[1]{\textcolor{blue}{[VB:#1]}}
    \newcommand{\akhilnote}[1]{\textcolor{green}{[AC:#1]}}
    \newcommand{\tristannote}[1]{\textcolor{cyan}{[ZY:#1]}}
    \newcommand{\aninote}[1]{\textcolor{green}{[AV:#1]}}
    \newcommand{\naveennote}[1]{\textcolor{blue}{[NV:#1]}}
    \newcommand{\wenxinnote}[1]{\textcolor{cyan}{[WX:#1]}}
    \newcommand{\jacknote}[1]{\textcolor{cyan}{[JL:#1]}}
    \newcommand{\kruthinote}[1]{\textcolor{cyan}{[KK:#1]}}
    \newcommand{\thrishnanote}[1]{\textcolor{blue}{[TB:#1]}}
    
    \newcommand{\gnote}[1]{\textcolor{magenta}{[GKT:#1]}}
    \newcommand{\jnote}[1]{\textcolor{olive}{[JCD:#1]}}
    \newcommand{\ynote}[1]{\textcolor{red}{[YHL:#1]}}
\else
    \newcommand{\vishnunote}[1]{}
    \newcommand{\akhilnote}[1]{}
    \newcommand{\tristannote}[1]{}
    \newcommand{\aninote}[1]{}
    \newcommand{\naveennote}[1]{}
    \newcommand{\wenxinnote}[1]{}
    \newcommand{\jacknote}[1]{}
    \newcommand{\kruthinote}[1]{}
    \newcommand{\thrishnanote}[1]{}
    
    \newcommand{\gnote}[1]{}
    \newcommand{\ynote}[1]{}
    \newcommand{\jnote}[1]{}
    \newcommand{\delete}[1]{}
\fi

\usepackage{authblk}
\title{More than one Author with different Affiliations}
\author[1]{Vishnu Banna}
\author[1]{Akhil Chinnakotla}
\author[1]{Zhengxin Yan}
\author[1]{Anirudh Vegesana}
\author[1]{Naveen Vivek}
\author[1]{Kruthi Krishnappa}
\author[1]{Wenxin Jiang}
\author[1]{Yung-Hsiang Lu}
\author[2]{George K. Thiruvathukal*}
\author[1]{James C. Davis\thanks{Direct correspondence to \url{davisjam@purdue.edu} and \url{gkt@cs.luc.edu}.}}

\affil[1]{Department of Electrical \& Computer Engineering, Purdue University} 
\affil[2]{Department of Computer Science, Loyola University Chicago} 

\date{}

\begin{document}

\title{An Experience Report on Machine Learning Reproducibility: Guidance for Practitioners and TensorFlow Model Garden Contributors}



\maketitle

\begin{abstract}

Machine learning techniques are becoming a fundamental tool for scientific and engineering progress. These techniques are applied in contexts as diverse as astronomy and spam filtering.
However, correctly applying these techniques requires careful engineering.
Much attention has been paid to the technical potential; relatively little attention has been paid to the software engineering process required to bring research-based machine learning techniques into practical utility.
Technology companies have supported the engineering community through machine learning frameworks such as \TF and PyTorch, but the details of how to engineer complex machine learning models in these frameworks have remained hidden.

To promote best practices within the engineering community, academic institutions and Google have partnered to launch a \SIG on Machine Learning Models (SIGMODELS) whose goal is to develop exemplary implementations of prominent machine learning models in community locations such as the \TFMG (TFMG).
The purpose of this report is to define a process for reproducing a state-of-the-art machine learning model at a level of quality suitable for inclusion in the TFMG.
We define the engineering process and elaborate on each step, from paper analysis to model release. We report on our experiences implementing the YOLO model family with a team of 26 student researchers, share the tools we  developed, and describe the lessons we learned along the way.

\end{abstract}
\section{Introduction}

\begin{figure*}[h!]
  \centering
    \includegraphics[width=\textwidth]{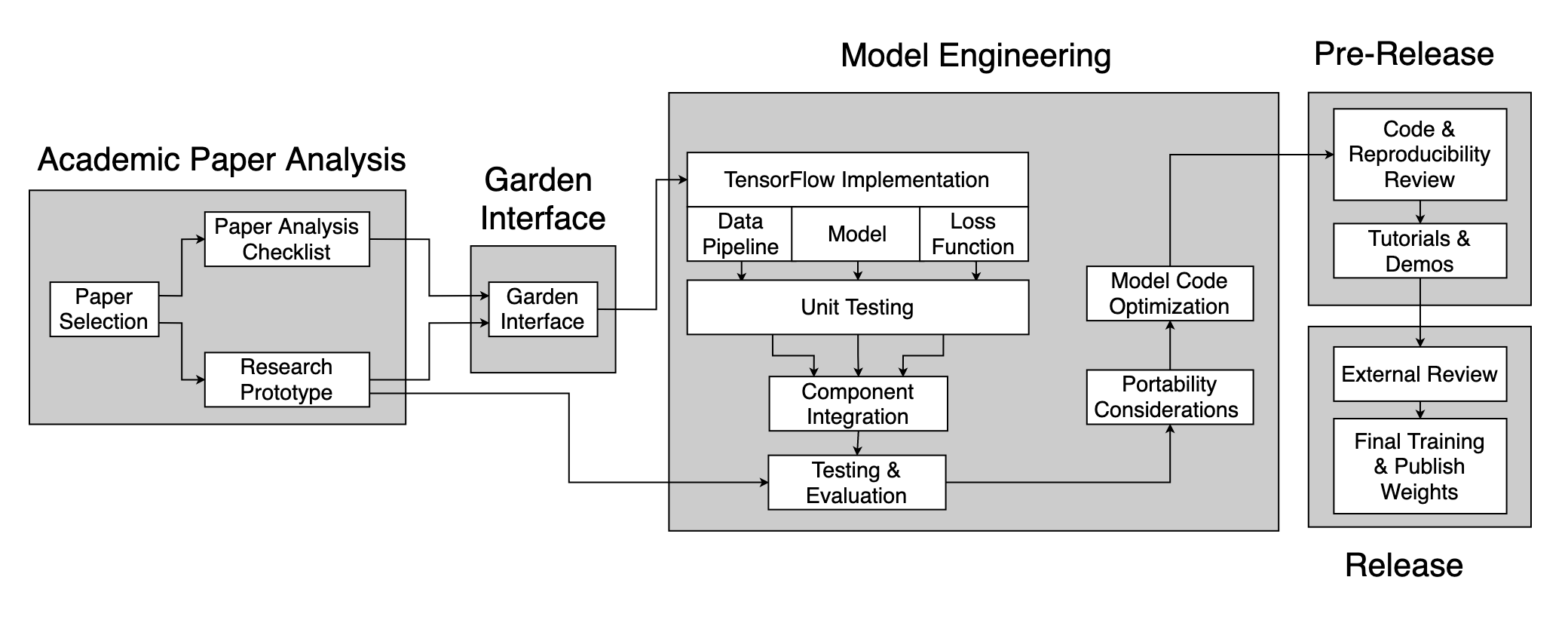}
  \caption{Our Model Engineering Process for the YOLO model family. Each phase of the project is described in detail and examples are taken from the YOLO exemplar.
  }
  \label{fig:paper}
\end{figure*}

Successfully reproducing a machine learning (ML) model, \ie implementing and contextualizing it, is a major problem facing the \ML community.
This process is followed in both the scientific and engineering communities, to compare against prior work and to transition research results into practice.
ML architectures are prototyped by researchers, but to bring the model to practice we need to increase the level of standardization of ML software.
Although an ML architecture may have several implementations, the lack of standardization makes it challenging to find, test, customize, and evaluate an existing implementation~\cite{Gundersen2018ReproducibleAI, Gundersen2018ReproducibilityinAI}.
These tasks require an engineer to combine libraries, reformat data sets, and debug third-party code.
This process is time consuming, error prone, and not resilient to component evolution.
An exemplary implementation of an ML model in a common framework, such as TensorFlow or PyTorch, makes this model more accessible to someone who seeks to extend or apply it.

The proposed solution is to provide a collection of \emph{Exemplars}.
Each exemplar is a reference implementation of an ML model that can be used and extended.
A collection of exemplars, implemented using the same framework and conventions, will be an engine to advance scientific and engineering progress. 
Such a collection would allow faster testing and customization of ML models, and would simplify development of new models from trustworthy components.
This concept is similar to the ``standard library'' in languages like C++~\cite{STL_robson2012} and Java~\cite{java_schildt2007}.
This task is too great for any one organization to complete, so it will require community contributions, similar to community package registries such as npm~\cite{npm}, Maven~\cite{Maven}, and DockerHub~\cite{dockerhub}.

In order to grow a collection of exemplars for the TensorFlow framework, dubbed the \emph{TensorFlow Model Garden},\footnote{\url{https://blog.tensorflow.org/2020/03/introducing-model-garden-for-tensorflow-2.html}.} a \SIG (SIG) has been established~\footnote{\url{https://github.com/tensorflow/community/blob/master/sigs/models/CHARTER.md}} with the objective of identifying and implementing state-of-the-art ML models in a consistent format~\cite{tensorflowmodelgarden2020}.
As a proof of concept, our pilot team at Purdue University is finalizing an exemplary implementation of several models from the YOLO family~\cite{redmon2016yolov1, redmon2017yolov2, redmon2018yolov3, bochkovskiy2020yolov4}.
This will soon be added to the \MG.

In this document, our goal is to describe the ML reproducibility process involved in engineering an exemplary implementation.
We seek to \emph{lead by example} and \emph{inform} SIG members and the broader ML community about the process and effective engineering practices.
\cref{fig:paper} depicts the reproducibility process that our team proposed and followed throughout the project.

Our contributions are: 
\begin{enumerate}
    \item We detail our process and criteria for selecting a TFMG paper (\cref{academic-paper}).
    \item We describe our engineering process, coupled with \TF code examples (\cref{model-eng}).
    \item We walk through the steps required to contribute to the \TFMG (\cref{pre-release} and~\cref{release}).
    \item We explain effective engineering practices to facilitate this process (\cref{effective-eng}).
\end{enumerate}

\textbf{Outline of Report} We begin with background material on engineering reproducible machine learning systems in \cref{sec:background}. Then we describe the engineering process in detail, along with code examples and instances from YOLO in \cref{sec:overview}-\cref{release}. We close by discussing engineering practices that will help novices establish engineering teams in the future in \cref{effective-eng}.
\section{Background and Related Work} \label{sec:background}
There are currently not many best practices defined with respect to reproducing an \ML model.
This makes \rep a significant problem within the \ML community. Though some platforms offer pre-trained \ML models, there is still a lack of knowledge in the fields of best engineering practices and \rep in this domain.

\subsection{Best Practices}

Deep learning is an emerging field whose probabilistic nature and underlying paradigm differs from traditional software~\cite{Karpathy2017Software2.0X}.
The engineering community has begun to adopt \DL, but lacks long-term experience in appropriate engineering methods~\cite{Lecun2015}.
Previous works have begun to enumerate the challenges and directions of best practice for \DL systems and applications.

Major technology companies have shared different kinds of studies on machine learning best practice, \eg Google~\cite{Breck2017aRubricforMLProductionReadinessandTechnicalDebtReduction}, Microsoft~\cite{AmershiMicrosoft2019SEforML:casestudy}, and SAP~\cite{Rahman2019MLSEinPractice}. Google~\cite{Breck2017aRubricforMLProductionReadinessandTechnicalDebtReduction} and Microsoft~\cite{AmershiMicrosoft2019SEforML:casestudy} provide high-level guidelines on the \ML engineering process. Breck \etal from Google present a rubric with 28 tests and monitoring practices to improve ML production readiness and reduce technical debt~\cite{Breck2017aRubricforMLProductionReadinessandTechnicalDebtReduction}. 
Rahman \etal present a case study of \ML engineering at SAP~\cite{Rahman2019MLSEinPractice}.
They discuss the challenges in software engineering, machine learning, and industry-academia collaboration and specifically point out the demand for a consolidated set of guidelines and recommendations for ML applications.
Unlike the guidance on high-level architectures and organizational processes shared by Google and Microsoft, we focus on lower-level engineering and programming patterns, and specifically focus on reproducing an existing model rather than building a new one.
New models will continue to be identified.
Our interest is in transitioning research results into practice by realizing known models in exemplary implementations.

Beyond industry leaders describing their practices, software engineering researchers have studied ML engineering practices in open-source software and in a wider range of firms~\cite{Zhang2019empiricalstudyofcommonchallengesindevelopingDLApplications, Zhang2019SEPracticeintheDevelopmentofDLapplications,Serban2020AdoptionandeffectsofSEBPinML,Washizaki2019a}. 
Zhang \etal report on the engineering life cycle of \DL applications~\cite{Zhang2019SEPracticeintheDevelopmentofDLapplications}. 
They indicate that validating \DL software is an open problem.
Based on their findings, they give suggestions to both practitioners and researchers, such as using well-known frameworks, improving the robustness of \DL applications, and adopting new fault localization tools.
Serban \etal conduct a survey, quantify best practice adoption, and indicate a positive correlation between best practices and software quality~\cite{Serban2020AdoptionandeffectsofSEBPinML}.
These works demonstrate critical needs for the proper management of different kinds of components in the engineering process and a significant role of testing and debugging tools and strategies. To deal with these demands, we provide a guideline on how to manage components in the \ML model engineering process, as well as a description of how to test and debug these models.

Our work thus fills a gap in both industry and academic knowledge.
We provide a novel perspective on ML best practices: an experience report by an engineering team itself, identifying a detailed engineering process and lessons learned.
Our specific focus is on ``deep learning'' software (\ie a non-trivial neural network-based model), but some of our findings will generalize to other forms of ML.

\subsection{Reproducibility}

The reproducibility of machine learning research is of growing importance, due both to the increasing number of sensitive applications, and the number of distinct \ML frameworks in which to implement models. The rapid increase of data, algorithms, and computation has resulted in the quick development of the machine learning process. As a result, the \ML community is paying more attention to the \rep of machine learning research papers in recent years~\cite{NeurIPS2019Reproducibility, NeurIPS2020Reproducibility}. Previous works have indicated the need for detailed guidance for engineers and researchers~\cite{Gundersen2018ReproducibilityinAI, Gundersen2018ReproducibleAI, Tatman2018AResearch}.

The significance of \rep is emphasized in the machine learning domain among researchers and practitioners. Pineau pointed out three needed characteristics for machine learning software: \rep, reusability, and robustness~\cite{Pineau2020}. She demonstrates that machine learning algorithms are usually unstable and difficult to reproduce in which case they proposes a need for proper documentation of the necessary information.
Pham \etal study model performance variance across \DL systems\cite{Pham2020TrainingDLSWsystemsAnalysisofVarariancee}. To combat this variance, they suggest careful consideration of training variance as well as the transparency of their research works, which can thus improve \rep of the works. Our work supports their suggestions. Moreover, we aim to provide detailed guidance for complicated works on reproducing existing models/algorithms or re-implementing the results of a published paper.

Machine learning research conferences have been trying to ensure that published results are reliable and reproducible for the last few years. In 2019, the Neural Information Processing Systems (NeurIPS) conference added a mandatory \rep checklist to their code submission policy, to promote the importance of \rep in future research~\cite{NeurIPS2019Reproducibility, NeurIPS2020Reproducibility}. 
Researchers are encouraged to re-implement parts of a paper and produce a \rep report which is similar to our work. However, the process and expert knowledge are implicit in the NeurIPS context which means they are not helpful enough for engineering practitioners.
The process we have outlined not only provides what we did but also indicates what and how should we do in the future. We are communicating this expert knowledge to both academic and engineering domains.

Our work provides detailed guidance for engineers who would like to reproduce academic research on \DL architectures for use in practice.
We make implicit academic knowledge explicit, and illustrate engineering processes.

\subsection{Towards Exemplary Deep Learning}

There are several preliminary attempts at developing exemplar collections in the \ML community including \TFMG, \eg \TF Hub~\cite{TensorFlowHub}, Torchvision~\cite{Torchvision}, and some models advertised via Kaggle~\cite{Kaggle}.
\TF Hub consists of different pre-trained models and datasets from across the \TF ecosystem~\cite{TensorFlowHub}.
Torchvision, integrated in Pytorch, collects popular datasets, models, and common image transformations for computer vision~\cite{Torchvision}.
Kaggle is a platform for users to upload their own trained models which are able to be downloaded and reimplemented by others.
Similar to these efforts, \TFMG is a repository of machine learning models and datasets built with \TF's high-level APIs~\cite{TensorFlowModels}.

A detailed guideline for engineering process of \ML models is valuable to both engineers and researchers.
In contrast, existing documentation is high-level.
For example, the Pytorch Contribution Guide~\cite{PytorchGuideline} indicates what engineers should consider in the engineering process of an open source project. This guide is approximately two pages long. In contrast, our work provides a more detailed guideline from paper selection to model engineering, as well as testing and releasing in \TFMG, which are not treated in the Pytorch document.
Although some of our technical advice is specific to TensorFlow, our engineering process and tools will generalize to other \ML frameworks.

\subsection{Exemplars: Research Prototypes Are Necessary, But Not Sufficient} \label{prot}

Research papers describing machine learning algorithms may lack the details required for complete model reimplementation~\cite{NeurIPS2019Reproducibility, NeurIPS2020Reproducibility}.
Hence, access to an actual implementation (\eg the research prototype, also known as the \emph{primary implementation}),  can prove beneficial for reference.
For example, Papers with Code~\cite{PapersWithCode} is a database of ML models with existing primary and secondary implementations using various libraries, including accuracy values and other associated metrics.

Although existing implementations are helpful, an \textbf{\emph{exemplary implementation}} is distinguished by its documentation, level of testing, and modular design to promote re-use and adaptation.
An exemplar is engineered, not programmed.
\section{Overview} \label{sec:overview}
The main objective of this document is to provide scientists and engineers members with an outline of the \ML reproducibility process, shown in \cref{fig:paper}.
Throughout the document, engineering best practices are highlighted, which are based on our model engineering experience. The process is divided into 5 main stages:
\begin{enumerate}
    \item \textbf{Academic Paper Analysis (\cref{academic-paper}):} We select a research paper for reimplementation and analyze it for key details.
    \item \textbf{Understanding the Garden Interface (\cref{garden-interface}):} To promote standardization, our exemplar implementation is constrained to follow the conventions of the TensorFlow Model Garden.
    \item \textbf{Model Engineering (\cref{model-eng}):} We implement the model using \TF, drawing on both the research paper and its original implementation.
    \item \textbf{Pre-Release (\cref{pre-release}):} We describe our internal review process, including testing, documentation, and evaluation to verify that the implementation is exemplary. 
    \item \textbf{Final Release (\cref{release}):} Our exemplar is reviewed by the \MG community. After all code is finalized, the model and trained weights are published.
\end{enumerate}
\section{Academic Paper Analysis} \label{academic-paper}
As illustrated in~\cref{fig:paper}, the first step in the ML reproducibility process is to select and comprehend the desired ML model. This section summarizes the relevant information to extract, such as the model structure and key components.
The information is presented in a checklist at the end of this section.

\subsection{Paper Selection}

SIGMODELS has a prioritized list of models that can be reimplemented and published in the \MG.\footnote{As of 28 June 2021, this list is maintained on the TensorFlow GitHub repository, at \url{https://github.com/tensorflow/models/issues/8709} and via the label ``help wanted:paper implementation''.}
The repository is also open to accepting exemplars from new research papers and they satisfy the criteria mentioned in the Research Paper Code Contribution guidelines.\footnote{See \url{https://github.com/tensorflow/models/wiki/Research-paper-code-contribution}.}

\subsubsection{Measuring Paper Difficulty} \label{model-difficulty}
Estimating paper difficulty will help your engineering team decide which paper to choose based on machine learning expertise within the group. The primary metrics we use to estimate the difficulty of reproducing a paper's model in \TF are:

\begin{itemize}
    \item \textbf{Architecture Size}: Smaller, simpler architectures are easier to implement. Size can be estimated using the lines of code and number of layers.
    \item \textbf{Custom operations}: TensorFlow includes some built-in operations. Custom operations must be implemented from scratch. This cost can be estimated by checking the operations used in the model, and comparing to the set available in TensorFlow. More details are in \cref{appendix-b}.
    \item \textbf{Existing implementations}: A primary or secondary implementation can clarify details not provided in the research paper. Unofficial implementations provide more opportunities for comparison and inspiration, particularly if the primary implementation does not use the desired \ML framework (\eg \TF).
    This can be measured in terms of number of official and unofficial implementations in existence, \eg with reference to a resource like Papers With Code~\cite{PapersWithCode}.
\end{itemize}

A difficult \ML paper will have
  a large architecture,
  more custom operations,
  and
  few existing implementations.

\subsubsection{Selection Advice}
It is recommended to select a project based on prior experience and expertise. If there is no existing knowledge related to the project, time-consuming preliminary research will be required. If the engineering team is new, it is advised to select one research paper, and add more as time progresses based on team size. Model difficulty can be measured using the metrics provided in \cref{model-difficulty} and should be adjusted based on the team's expertise in the field of \ML. If the team has a specialization, then the purpose or type of the model will also affect the paper selection process. For example, our team focuses on \CV and image processing and thus, the projects we select will be related to the \TF Vision API. 

\subsubsection{\TF Restrictions} \label{tf-restrictions}
Before finalizing your paper choice, analyze the feasibility of the project by checking whether the operations in this paper can be replicated in \TF using existing or custom operations.
Not all operations are possible in \TF.
Study the paper and existing implementation(s) to determine whether all functionality can be implemented.

For example, our team rejected the \emph{Momentum Contrast for Unsupervised Visual Representation Learning} (MOCO) project because the functionality of a shuffle batch norm was not possible in \TF at the time.
When you identify such a limitation in \TF, is good practice to create an issue on the official \TF GitHub repository describing the missing operation or functionality.
We did so, and the \TF team introduced support for a shuffle batch norm in TensorFlow v2.4. 

\subsection{Paper Analysis Checklist}

To successfully engineer an exemplary implementation of the model you selected, you will need to understand its architecture and context.
Here we present our paper analysis checklist, to help you extract information from the authors' description of their model. 
The checklist provides insight regarding the potential difficulties and considerations during model reimplementation.
Since machine learning models vary, you may need to tailor it to the model you are re-implementing. 

To illustrate the use of this checklist, we applied it to the YOLO v3 paper~\cite{redmon2018yolov3}.
This helped us identify several potential difficulties and considerations during our YOLO replication.

For example: 
\begin{itemize}
    \item \textbf{Framework}: The primary implementation uses a different framework, which will need to be addressed as mentioned in \cref{prot}. 
    \item \textbf{Custom Layers}: Some of the layers required for YOLO do not exist within the \TF API. These custom layers must be implemented, following conventions so that other exemplars can use them.
\end{itemize}

The checklist follows.
The result of applying the checklist to the YOLO v3 paper is indicated at the end of each checklist item, denoted with \emph{YOLO example}.

\vspace{10pt}

\begin{tcolorbox}[enhanced,drop shadow]
    {\Large General Checks}
    \begin{description}
    	\item[Model Purpose] The model development approach depends on the type of model being implemented and its primary function. Deep Learning code performs a variety of tasks. Typical examples are object detection (OD) and natural language processing (NLP). \emph{YOLO example: Detect Objects, Classify Objects.}
    	\item[Code Availability] The existence of an official or accurate unofficial implementation can serve as a reference during the model engineering process. \emph{YOLO example: Original Darknet Repository.}
    	\item[Language/Framework/Libraries Used] A model implementation will vary based on the programming language and the \ML library used. Some operations will directly correspond to language or framework functions, while others will require custom operations. \emph{YOLO example: Darknet C Library.}
    	\item [Networks Referenced] If another neural network is referenced, you may need to refer to it while reconstructing your exemplar. \emph{YOLO example: ResNet, YOLO v2, Faster RCNN, RetinaNet.}
    \end{description}
\end{tcolorbox}

\vspace{10pt}

\begin{tcolorbox}[enhanced,drop shadow]
    {\Large Model and Design Checks}
	\begin{description}
		\item [Model Architecture] The model cannot be built without knowing its architecture,  i.e the type, size and connection of layers. \emph{YOLO example: Architecture provided.}
		\item [Model Sub-Networks] Sub-networks are sub-structures, such as a backbone or decoder, that can be built independently because they serve a discrete purpose. These elements can be built in parallel during the model engineering process. \emph{YOLO example: Backbone, Classification Head.}
		\item [Model Building Blocks] Building blocks are a set of layers or components that are repeatedly seen in a particular order within the model architecture. Having well-defined building blocks will ensure  components is being reused whenever possible, with no  redundant code segments. \emph{YOLO example: DarkResNet Block, Routing Layers, DetectionRouteProcessing, DarkBlock.}
		\item [Custom Layers] TensorFlow has predefined layers, but most Machine  Learning models will require custom layers that perform a  specific function. Once a custom layer is built, it can be used  anywhere within the Model Garden. \emph{YOLO example: DarkConv.}
		\item [Loss Functions] The loss function is used to train an \ML model. The model parameters are tuned until the loss function takes on small-enough values. \emph{YOLO example: Loss equation is provided.}
		\item [Output Structure] Knowing the final output structure will help during the unit and   differential testing phases of model development. \emph{YOLO example: Architecture is provided.}
	\end{description}
\end{tcolorbox}

\vspace{10pt}

\begin{tcolorbox}[enhanced,drop shadow]
    {\Large Training and Evaluation Checks}
	\begin{description}
		\item [Dataset Used] Without using the same dataset as the research paper, it is  not possible to reproduce the results obtained in the original implementation. Most modern models are evaluated on well-known datasets. \emph{YOLO example: COCO dataset.}
		\item [Pre-Processing Functions] Pre-processing functions are used to format  and augment the dataset before feeding input to the model. \emph{YOLO example: Random jitter, random crop, random zoom.}
		\item [Output Processing Functions] Once the model has processed the input, the output may be adjusted. \emph{YOLO example: No post-processing.}
		\item [Testing and Target Metrics] Testing Metrics are metrics used to evaluate a machine learning model. Some metrics are built into TensorFlow.  Others must be custom-made. Target Metrics  are the metric values obtained in the original implementation, and will be provided in the research paper. The closer the  exemplar's performance metrics are to the original one, the more precise the exemplar is. \emph{YOLO example: Several measures of average precision --- $AP_{50}$\, $AP_{75}$, $AP_S$, $AP_M$, $AP_L$.}
		\item [Training Steps] A model may have special steps to follow during the training process. \emph{YOLO example: The input format and training steps are provided.}
	\end{description}

\end{tcolorbox}


\section{Interfacing with the Exemplar Conventions} \label{garden-interface}

This following section discusses the organization and interfacing requirements for \TFMG exemplars. All exemplars in the \MG follow the organizational structure depicted in \cref{garden-table}. 

{
\renewcommand{\arraystretch}{1.2} 
\begin{table}[h!]
    \centering
    \caption{The organizational structure of \TFMG exemplars, grouped by whether each directory is (a) Core; (b) Supporting; or (c) Optional.}
    \vspace{0.25cm}
    \begin{tabular}{llp{11cm}}
    \hline
    \toprule
      \textbf{Folder} & \textbf{Required} &        \textbf{Description}     \\
    \toprule
    dataloaders & Yes & Decoders and parsers for your data pipeline. \\
    modeling    & Yes & Model and the building  blocks. \\
    losses      & Yes & Loss function. \\
    \midrule
    common      & Yes & Registry imports. The tasks and configs need to be registered before execution. \\
    configs     & Yes & The \code{config} files for the task class to train and evaluate  the model. \\
    ops         & Yes & Operations: utility functions used by the data pipeline, loss function and modeling. \\
    tasks       & Yes & Tasks for running the model. Tasks are essentially the main driver for training and evaluating the model. \\
    \midrule
    utils       & No  & Utility functions for external resources, \eg downloading weights from the primary implementation's repository, datasets from external sources, and the test cases for these functions. \\
    demos       & No  & Files needed to create a Jupyter Notebook/Google Colab demo of the model. \\
    \bottomrule
    \end{tabular}
    \label{garden-table}
\end{table}
}

\subsection{The Task Structure} \label{task-struct}
In order to unify and automate model testing, the \TFMG has a custom training library, \code{orbit}. This creates a unified structure that any contributor could implement quickly to test a model, while also preserving the functionality and autonomy of the Garden Trainer. The basic unit of a \TFMG is called the Task, which can be found in the \code{official.core.base\_task} file. This class serves as an interface allowing you to use the \code{train.py} file to automatically train any model using the command line.

The Task contains the general interface for a model in the \TFMG. It is responsible for building and unifying the main components (model, loss function, and data pipeline) and applying the appropriate evaluation metric(s). It also should contain the \code{train\_step} and \code{validation\_step} methods required to train and validate model performance. These should be tailored to fit the needs of a specific implementation. Further details regarding component integration are discussed in \cref{component-integration}.
 
\subsection{Configuration}
In order to configure a task, each implementation will require the addition of a set of dataclass configurations that inherit from those found in \code{offical.core.config\_definitions}. In addition to the configuration files, you should include a method named \code{experiment} that predefines all the configuration parameters and serves as a default model state if a given parameter is not in the operating configuration file. This will allow configuration files to remain concise while also preserving the model's essential functionality. The configurations and the Tasks are designed to come together into an operational model that can be manually configured via the input of a model configuration, or automatically configured using the trainer and a configuration file. Given that the model operates using this dual functionality, the task will not have any class parameters other than \code{self.task\_config} used to hold the input configuration dataclasses.
\section{Engineering An Exemplar} \label{model-eng}

As seen in~\cref{fig:paper} and~\cref{garden-table}, all \DL models, regardless of their type or purpose, can be broken down into three main components: the Data Pipeline (\code{dataloaders}), Loss Function (\code{losses}) and the Model itself (\code{modeling}). These components can and should be developed and tested independently. 

\subsection{The Extract-Transform-Load Data Pipeline}
\begin{figure}[h!]
    \centering
\includegraphics[scale=.6]{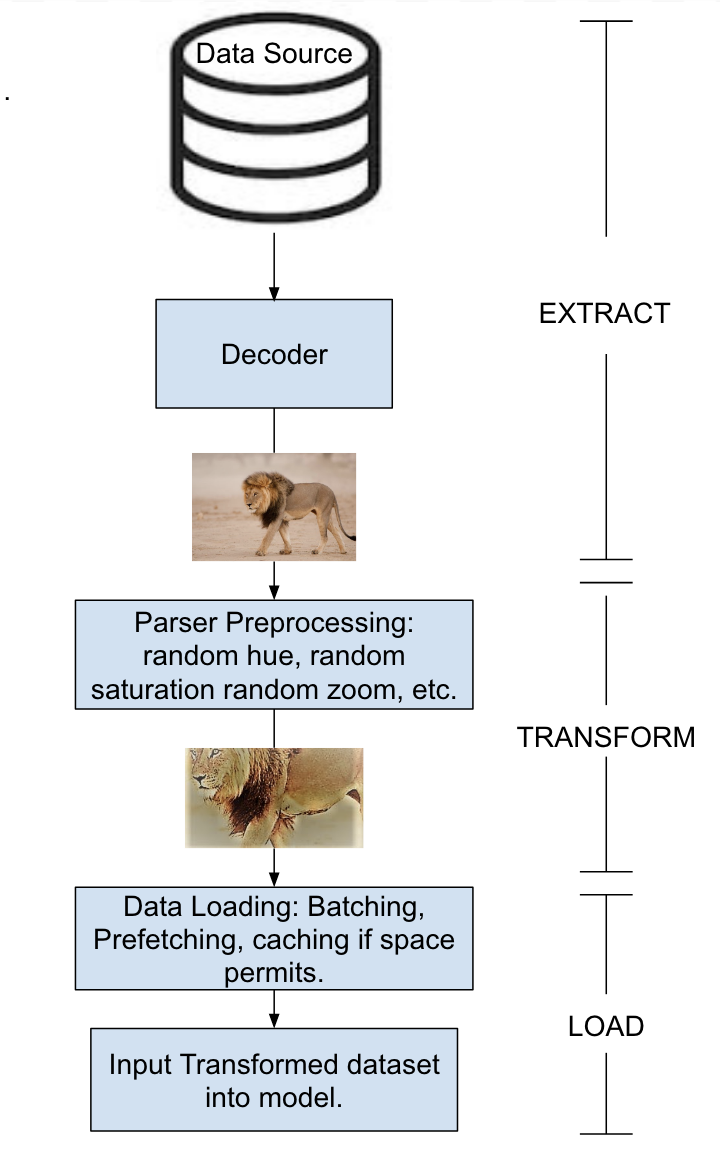}
    \caption{
    Components of a general Data Pipeline, along with the ETL architecture. This pipeline is specific to YOLO.
    }
    \label{fig:data-pipeline}
\end{figure}

\cref{fig:data-pipeline} depicts the common structure of the Data Pipeline, as instantiated in our YOLO exemplar.
The \emph{Data Pipeline} extracts raw data such as pictures from a storage source (\emph{decoder}), transforms the data into the required input format (\emph{parser}), and loads the transformed data into the GPU (\emph{loader}). Its speed matters: the Data Pipeline can be a bottleneck for training and inference. 

In the \TFMG, the Extract-Transform-Load (ETL) architecture~\cite{Wiki:ETL} is the preferred approach for the Data Pipeline.\footnote{See: ~\url{https://github.com/tensorflow/docs/blob/master/site/en/r1/guide/performance/datasets.md}.}
This modular design promotes reuse in different \TFMG exemplars.
Different models may share datasets (Extract), augmentations (Transform), or final stages (Load).

Each exemplar's decoder and parser should be inherited from the \code{decoder} and \code{parser} classes of the \TFMG.\footnote{As of 28 June 2021, these classes are defined in \url{https://github.com/tensorflow/models/tree/master/official/vision/beta/dataloaders}.}
The loader serves as an interface between the data pipeline and the model, and will vary based on architecture.

\subsubsection{ETL -- Extract}
The extraction component, called the decoder, converts the raw dataset to a format that is compatible the rest of the Data Pipeline. It is imperative before developing the decoder to determine the raw data format structure. After this, the data should serialized into a format that will be compatible with the data handling functions.

TensorFlow’s data handling functions are found in \code{tensorflow.data}.
This API interfaces with and is optimized for \TF's own file format, known as \code{TFRecord}, which is a serialized sequence of binary records. The \code{tensorflow.train} API and \code{tensorflow.io} API convert raw inputs to \code{TFRecord}, which can later be loaded in as a \code{tensorflow.dataset} object.
After converting and standardizing the raw input into a \code{TFRecord}, it is ready to be passed onto the transformation component. 

Note that there should be one decoder class for a specific dataset and this should be inherited from the \MG's \code{decoder} class.
This promotes two kinds of standardization: among decoders within the project, and among inputs going into the transformation component of the data pipeline.

\subsubsection{ETL -- Transform}
The transformation component is called the parser.
It handles preprocessing, normalization, and formatting of the input features and labels within the dataset to fit the model input format. The parser should be a class that contains two data handling methods, one for training and another for evaluation. There should also be another method that returns one of these methods based on the activity being done. The returned method should contain all of the necessary data handling functions to prevent incompatible library dependencies and optimize performance.

The function returned by the parser can be mapped onto the dataset, but only modifies the features and labels that are loaded from a persistent storage source into system memory and not the dataset itself. This will allow the model to begin training faster as only the first batch will be required to be processed and using prefetching, data can then be processed on the CPU while the model trains on the GPU, which will reduces both processing units' idle time. Another way to optimize the efficiency of the transformation component is by parallelizing the mapped functions which reduces the amount of time it will take the CPU to execute the mapped function over the batch. Note that there should be a parser depending on the type of training and phases of that training. Similar to the decoder,
there should be one parser class for a specific dataset and this should be abstracted from the \MG's \code{decoder} class.

\subsubsection{ETL -- Load}
The loading component is responsible for loading the dataset into the model for training.
This component also applies dataset augmentations such as batching, prefetching, caching, shuffling, and interleaving.
The dataset augmentations are primarily determined by the size of the dataset such as how the dataset interacts with system memory.
The loading component is located in the Task structure in the method \code{build\_inputs()},
This method instantiates the decoder and parser, and the entire loading component, \code{input\_reader}.

If a dataset fits in the available system memory then it can be cached.
Otherwise, the dataset must be processed in batches.  If the dataset is larger than system memory but caching is attempted, the process or the machine may crash. Shuffling and interleaving should be done with the understanding that the larger the dataset, the longer the time it takes to start training for each epoch.

\subsection{Model}

Generally speaking, a model is an ordering of operations parameterized by weights that control its behavior.
This structure uses input data to accomplish a goal, \eg YOLO's goal is object detection.
To structure the model, operations and weights are grouped into layers, which are themselves combined into blocks.
The relationship between these model elements is shown in \cref{fig:model_structure} and explained in \cref{appendix-a}.
Each layer and block can be designed and tested individually, and then integrated into the full model structure prior to training.

\begin{figure*}[h!]
    \centering
    \resizebox{\textwidth}{!}{\includegraphics{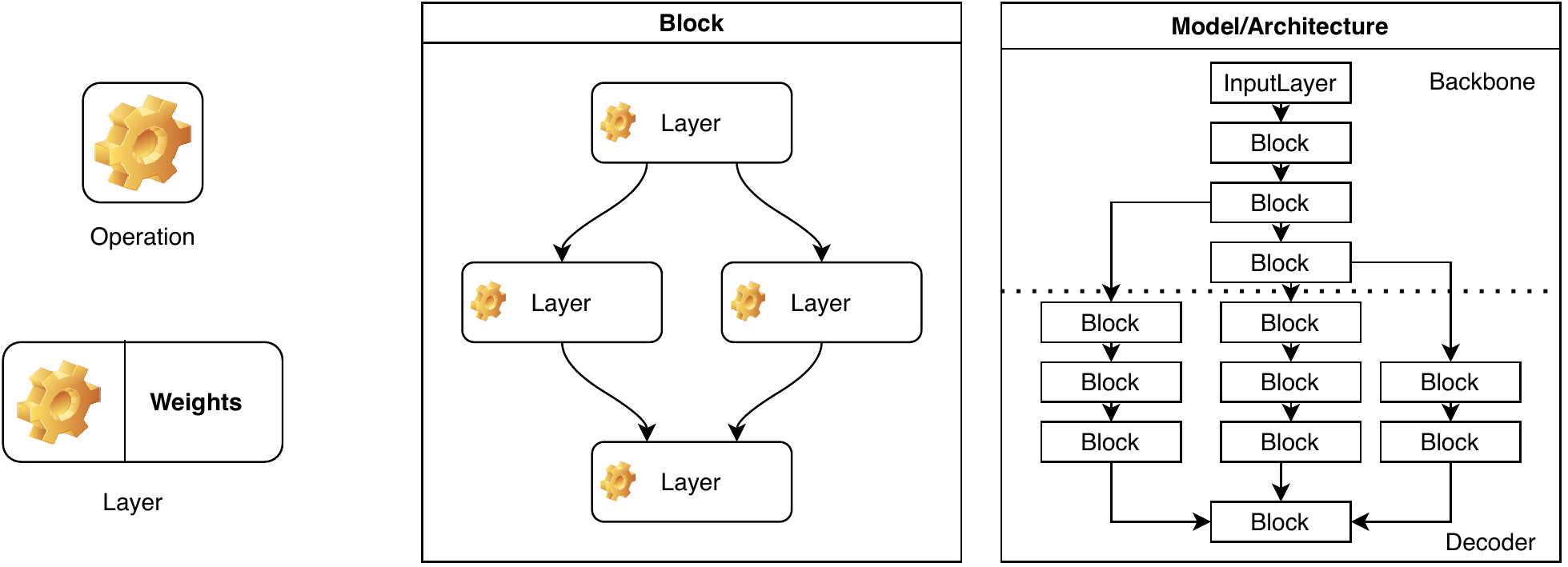}}
    \caption{
    Relationship between model components, from simple operations to the full architecture. 
    }
    \label{fig:model_structure}
\end{figure*}

The architecture of a \ML model varies based on its purpose. For computer vision, models have three main components: backbone, decoder, and head.
The backbone is the combination of layers responsible for feature extraction. The backbone is usually a large number of sequential layers and the majority of the model training process is to adjust the backbone parameters.
A model decoder, not to be confused with the data pipeline decoder, reasons using the features extracted by the backbone.
It usually has fewer layers than the backbone.
Lastly, the head, often the smallest element, is responsible for the final model task (\eg image classification).

Many models re-use other models as modules, \eg combining one model for feature extraction with another model for inference from those features.
One of the values of a collection of exemplars is that it enables this type of model re-use.
Thus, while constructing the model, ensure that the structure mentioned in \cref{fig:model_structure} is followed.

Throughout the model development process, continually refer to the model architecture provided within the research paper.
Usually, the authors will describe architecture in detail, including the number of layers, size of each layer, how each layer is connected to other layers, and fine-grained optimizations such as parameter adjustment.
Identifying repeating layers and layer connections within the architecture can help create a ``building block'' to reduce the lines of code of the model itself 
Many model architectures will require at least one custom layer that does not exist in the \TF API.

\subsection{Loss Function}
The final model-specific component is the Loss Function.
The loss function is a function that measures the extent to which the model's performance matches the ideal behavior.
The loss function is used to train a model, adjusting the weights to reduce the loss.
The loss function is not used after the model is trained.
The loss function must be differentiable because it defines the gradient flow through the neural network by backpropagation. Therefore, the loss function must only use differentiable operations from the TensorFlow library, such as those found in the \code{tf.math}, \code{tf.nn}, \code{tf.keras.backend}, or the \code{tf.keras.losses} APIs.

Implementing a loss function can be more intricate than other components like the layers or the blocks.
When computing derivatives during training, TensorFlow must track variables, consuming memory and time.
Training is expensive, and thus the loss function is an optimization target.
Not all computations in the loss function need to be derived.
Thus, to keep the loss function and the overall footprint of the loss function in check, it is best to tell TensorFlow what to ignore when computing your derivatives. This can be accomplished by using the \code{tf.stop\_gradient} method. This operation will mark tensors so that the gradient tape will view them as a constant, and ignore any computation that lead to the creation of the flagged tensor.

Since it is a mathematical function, the loss function derivative can be manually computed and verified.
During our implementation of the YOLO exemplars, we found it necessary to manually compute several loss function derivatives in order to understand the causes of differences between the Darknet and \TF implementations.

\subsection{Testing \& Debugging}

During our exemplar implementation, we introduced and later repaired two types of bugs.
``Programmer bugs'' are logical errors that are not specific to \DL, \eg off-by-one errors.
``Framework bugs'' involved misunderstandings of the APIs of the \DL framework.
We found most of our programmer bugs early on (\eg they caused crashes), but did not find many framework bugs until later (\eg they degraded end-to-end model accuracy).

Each component of the implementation was tested individually (unit testing).
Then, after integration, the entire network was tested.
Throughout, we applied two main types of testing: differential testing and visual testing.
Differential testing compares the output of the original implementation to the re-implementation numerically, whereas in visual testing, it is the programmer's responsibility to use appropriate data visualization techniques and then manually compare the outputs of both models. Details of both testing methods are provided in \cref{diff-testing} and \cref{vis-testing}.

\subsubsection{Unit Testing} \label{unit-testing}
Unit Testing must be performed on each component using unit test cases. This will ensure that all components perform as expected prior to component integration. Thus, if there are any inaccuracy in the final implementation after component integration, we can focus on the component boundaries.\footnote{In \TF, unit test cases are made using \code{tf.test.TestCase}.}

\textbf{Unit Testing - Data Pipeline. }
Since the Data Pipeline is non-deterministic, the randomized variables must be fixed during the testing phase. Each pre-processing operation must be tested separately by passing a sample image through the original pipeline and your \TF pipeline. Each operation must be tested using differential testing and if this fails, apply visual testing. Since the Data Pipeline is the rate-determining step for the training process, the time consumption of the overall pipeline must be noted and compared with the original implementation. If it is significantly slower than the original one, then it must be modified by optimizing each operation.

\textbf{Unit Testing - Model.} 
We applied several types of unit tests to the model.

\begin{itemize}
    \item \emph{Passthrough tests} check whether data can pass through the model. These are similar to so-called ``sanity tests'' or ``smoke tests''. If they fail, they indicate that the model architecture is not connected properly, or that the output shape of some model layer is incorrect. The pass-through test can be applied to layers and building blocks, while the other tests require the full model.
    \item \emph{Serialization tests} confirm that the model is serializable and deserializable. This verifies the $\code{get\_config}$ and $\code{from\_config}$ functions of the model and verifies that the model can be saved and checkpointed.
    \item \emph{Gradient tests} calculate the model's gradient using a gradient tape and check that the output is differentiable on every input. Gradient tapes are used to store the operations and gradient values for automatic differentiation, which makes sure that only the necessary tensors are differentiated. If the model has a non-differential component, an element of the gradient will be $\code{None}$ and the input has no effect on the output. This likely means that the input is not used correctly in the model and there is a bug in the model.
    \item \emph{Fixed-weight tests} involve loading \emph{pre-trained} weights from another implementation of the model, printing the output of each layer, and comparing to the other implementation. This will help identify which particular layer is incorrect in the model. Model elements, namely blocks and layer, must be tested similar to the model. 
\end{itemize}

\textbf{Unit Testing - Loss Function. }
For testing the loss function, generate random tensors, feed them into both implementations (yours and the comparison implementation), and compare the output. The outputs may not be identical due to framework rounding for numerical types, so if the outputs are within rounding errors of one another (determined manually), then it can be concluded that the loss function is correct. Each component of the loss function can be tested separately. 

\subsubsection{Differential Testing} \label{diff-testing}
Differential Testing is the process of comparing two supposedly-identical systems by checking whether they are input-output compatible~\cite{McKeeman1998DifferentialTesting}.
Given the purpose and context of exemplar implementations, differential testing is a natural validation approach.

Differential testing can be applied to deterministic and probabilistic components, but the approach will vary. If the component is deterministic, then the same input can be passed in the original and reimplemented component and simply compared (within floating-point approximation).
If the component is probabilistic, then checkpoints can be run using the same weights, which will convert the process into a deterministic comparison.
Alternatively, probabilistic components can be checked for equality within a tolerance.

As recently demonstrated by Pham \etal~\cite{Pham2020TrainingDLSWsystemsAnalysisofVarariancee}, crossing deep learning framework boundaries introduces additional variance. 
They report that in two \DL frameworks, even with an identical training schedule applied to the same model, the trained weights may not match.
The accuracy can vary significantly, and thus, training should not be used as the only final testing method.
The original weights/checkpoint should be loaded into the network and then the corresponding outputs must be compared to the original implementation.
This will remove the non-deterministic process of training, and thus, will serve as an accurate testing method for model comparison.
We described these as ``fixed-weight tests''.

Due to differences between programming languages and deep learning frameworks, there are limits to the granularity at which differential testing can be applied.
Language and framework semantics may require implementing the same functionality in different ways.
In this case, the sub-steps of these different implementations cannot be compared using differential testing.
Therefore, unit testing complements differential testing.
At a coarse granularity, differential testing is valuable; at fine granularity, unit testing is needful.

\subsubsection{Visual Testing} \label{vis-testing}
We found that differential testing worked well in most cases, but that it was difficult to apply to all components.
For example, the overall Data Pipeline has complex probabilistic output whose properties are hard to measure for use in an automated comparison. 
In this case, we found Visual Testing useful.
We visually inspected the outputs of both pipelines on a sample input and looked for a noticeable difference between the two. In the case of YOLO, since the final output is based on bounding boxes, the resulting output will have a confidence level associated with each box. Ensure that the re-implementation output format matches the original implementation format, such as dimensions and color in the case of images.

Simple errors, such as image padding, can be detected via visual testing.
Incorrect padding can be noticed as soon as the image is printed --- these errors may be easier to ``see'' than to measure.
However, for more intricate checks, such as hit-maps, the data visualization must be modified, as a simple printed image will not be sufficient. Once the hit-map visualization is printed, it can be compared to the original implementation and bugs can be identified.
At some point a visual inspection becomes inadequate, and automated fuzzy comparisons are necessary.

\subsubsection{Fault Localization}
Based on what aspect of the model output is different from the expected output, the location of a fault can be narrowed down to one of the three main components of the implementation.
If the data augmentation outputs do not match the comparison implementation, then the bug can be localized to the Data Pipeline.
To test the model, use the \code{model.summary} function to obtain the number of parameters and compare each layer to the original implementation.
Finally, to test the loss function, use the same training schedule as the original implementation.
If the loss curves are dissimilar, the bug likely lies in the loss function. 

It took us several months to debug our exemplar implementation.
Bug localization was a bottleneck in the model engineering process.

\subsubsection{Debugging Practices}
One difficulty in testing an exemplar implementation is the cost (\eg cloud spend) required to train a model during the code-test-debug-fix cycle. From our experimentation, it is seen that 70\% of final results were obtained within 10-25\% of training, similar to an ``80-20 rule''. Thus, we advise that if the evaluation metric is not comparable to the original implementation after 25\% of the training, then your implementation will require debugging. This shortens the debugging cycle as the entire training process does not need to complete in order to identify an error.

During the debugging process, maintain consistent configurations and a log file of all the changes made to track results (provenance).
Debugging is time-consuming, so it must initially be applied towards components that will increase the evaluation metric for the implementation. For example, correcting the loss function will have a major impact on the weights of the model.

\subsection{Component Integration} \label{component-integration}

While assembling the required components, ensure that all model parts are consistent with one another. All components must be integrated with the Task Structure mentioned in \cref{task-struct}. Within the Task Structure, the data pipeline, model, loss function and metrics are built. If there are interfacing issues between components, these will arise while running the \code{task}. The \code{task} must handle the integration issues by making all formats uniform such conversion functions, such as the bounding boxes format in YOLO.

\subsubsection{Integration Issues}
Per~\cref{fig:paper}, we divided engineering work among sub-teams by component. Each team  developed and tested their components independently. Unit testing is discussed in~\cref{unit-testing} in conjunction with broader testing concerns. After unit testing, we combined the components. The integration process uncovered several issues. The main problems that we encountered during component integration were type errors and interface errors. Such issues are straight-forward to fix and will require minor modifications, such as casting or size adjustment, to the code-base. All integration issues are caught by the compiler and the error messages provided clearly indicate the type of inconsistency that has occurred.

\textbf{Type Errors.}
In \TF, several functions accept only certain types and thus, variable types must be carefully selected and cast when required. For example: if the data pipeline outputs \code{bfloat16}, but the model requires another type as input, then it must be cast at the interface.

\textbf{Pipeline-Model Errors.}
Throughout the model, shape consistency must be maintained so that there are no errors when performing tensor and matrix operations. For example: if a computer vision model requires a particular dimension image as an input to the first layer, but the data pipeline outputs a different dimension, an image resize or crop operation must be performed to properly interface the data pipeline and the model.

\textbf{Model-Loss Errors.}
Each entity in a \ML model can be encoded differently based on the engineer's judgment. For example, in object detection, a bounding box can be defined in terms of the center, height and width or two opposite corners. When integrating the loss function and the model, ensure that both components use the same attributes to define a bounding box.

\subsection{Evaluation}

Once the individual components are determined to be correct using individual testing (\cref{unit-testing}), and they have been integrated (\cref{component-integration}), the next step is to check if the entire model is functioning as expected.
The target metrics must be implemented correctly and then used for comparison.
If all target metrics match those seen in the original implementation, then it likely that the model is an accurate reimplementation.
However, recall that these metrics are aggregate summaries of the performance of the model.
There are cases where the metrics can match, but the model's output remains systematically incorrect.

\subsubsection{Metric Evaluation}
The metric that is being used to determine the training condition of the model must be the same metric that is maximized in the original paper. Some examples of the metrics are accuracy for classification models, Mean Squared Error (MSE) for regression models, Intersection over Union (IoU) for computer vision models, and Perplexity for natural language processing models. Before implementing a custom metric, check if the metric exists in the \code{TensorFlow.keras.metrics} library or the \TFMG. If the metric is not found in either of those locations, then custom metric implementation is necessary. For our YOLO model family, the metrics were already implemented in the \MG as \code{coco\_evaluator.py}. If a custom metric is needed, then inherit from \code{TensorFlow.keras.metrics.Metric} and use functions from \code{TensorFlow.Math} to do the necessary computations. Most state-of-the-art models use a metric that is already a built-in \TF metric or can be made into a custom one using \code{TensorFlow.Math} functions.

Comparing obtained metrics to the target metrics is done to determine the accuracy the reimplementation. If the metrics output is off at the early stage of training, tweak the learning rate or make other suitable modifications. For Object Detection, the main metric that YOLO uses is Mean Average Precision (mAP), which can be compared by using the same dataset as the original implementation.

\subsubsection{Unit Evaluation}
Metric evaluation will state whether the re-implementation has identical aggregate performance when compared to the original implementation. However, this does not imply that the models produce identical results. Using metric evaluation for model evaluation will provide a rough estimate for re-implementation accuracy, but more direct and involved testing is required. For example, in the case of YOLO, if the original implementation generates bounding boxes that are all 5\% too far left from the ground truth, and the re-implementation generates boxes that are all 5\% too far right from the ground truth, then the mAP metric will be identical for both, but it cannot be claimed that the model is accurately reproduced. 

Therefore, in addition to evaluating the model on a large dataset, the model should also be evaluated on a small set of inputs.
This is called Unit Evaluation: evaluating an individual input, \eg an image, and comparing it to the original implementation. For YOLO, when using unit evaluation, the bounding boxes can be manually compared. This process should be repeated for multiple images. If a trend is observed, such as in the case mentioned above, then the model must be adjusted accordingly. Unit evaluation is done using the help of visual testing, as described in \cref{vis-testing}.
This process can be automated, but we did not find it necessary to do so.

\subsection{Portability Considerations} \label{portability}
For an implementation to be published to the \TFMG, it must be capable of training and evaluation across the common classes of processors: \CPU (CPU), \GPU (GPU), \TPU (TPU); and in resource-poor environments like mobile devices (via TFLite). Until now, we have been focused on functionality, and not the portability of the implementation.

Training and evaluation on a TPU and a mobile device are the most restrictive, with two main constraints.
The first constraint is that these contexts support limited operations.\footnote{See, for example,~\url{https://cloud.google.com/tpu/docs/tensorflow-ops}.}
We have found the public documentation somewhat wanting in this regard.
Some operations are permitted, but only in certain configurations and situations.
For example, at time of writing, TPU operations require \code{bfloat16} type for input parameters, and the \code{upsampling} operation behaves correctly during forward propagation of the network but not backward on the TPU.
The other constraint is the shape of the tensor must be known at each stage of the model. If the shape is unknown for a critical operation, then the process will throw an error, else it will take longer to train. The output shape for a particular layer must be calculated in advance for memory allocation purposes.

We advise you to try the operations that you plan to use early on to test for support.
In addition, run your completed model on both a TPU and a mobile device to check for a change in the model behavior as compared to the behavior on a CPU or GPU.
All models are capable of running on a CPU, as long as they compile correctly.
GPU has certain restrictions based on the computational power required for the operation, but all operations will function correctly with differing speeds.


\subsection{Model Code Optimization}
After the model meets the target accuracy metrics and is capable of training and running on a CPU, GPU, and TPU, the final step of the engineering process is to optimize the model speed. The model must be optimized without changing the output of the \ML model. An example of this would be replacing \code{for} loops with tensor algebra for parallel computation. There are two main aspects of model optimization: graph execution and model training.


\subsubsection{Graph Execution}
Graph execution time can be optimized by using existing TensorFlow function calls rather than custom operations, because these functions have been optimized.
Shape consistency must be maintained throughout the model and it must match the original implementation at each layer. Identity layers and blocks should be removed to reduce graph size and decrease number of operations and parameters. Memory occupied by the model graph must also be carefully allocated by removing bottlenecks and optimizing parameters based on the device.

\subsubsection{Model Training}
Based on the device being used for training, the model training time and efficiency will vary drastically. However, there are certain measures that can be taken to increase the training efficiency on average. The Data Pipeline operations have a strong effect on the training time and hence, the operations used in the Data Pipeline stage must be efficient and based on \TF built-in operations whenever possible. It is imperative to adjust the epochs and other parameters based on whether the training device is a CPU, GPU or TPU. Some of the Data Pipeline pre-processing operations take a long time on a CPU because they are computationally expensive. For example: in a previous iteration of the YOLO Data Pipeline, image blurring was included, which required the convolution operation. However, all the remaining operations did not require GPU resources and worked optimally on a CPU. Transitioning to a GPU for only one operation and then back to the CPU would be time-consuming.
This issue can be addressed by moving the blurring operation to the end of the Data Pipeline, and transitioning the data to the GPU \emph{before} performing blurring. Then, blurring is done on the GPU, and the training process can begin immediately after.

Thus, the order of operations can influence the model's training time.
Within the bounds of commutativity, operations may be shuffled as an optimization. 
We note that this breaks the modular design of the data pipeline; breaking module boundaries is a common effect of performance optimization.

\section{Pre-Release} \label{pre-release}
Once the model engineering has been completed, there are some essential steps before submitted the final model to be published in the \TFMG. The following section describes the internal review steps that our team took before external review.

\subsection{Code \& Reproducibility Review}

Lack of proper documentation in model repositories is a problem within the ML research community~\cite{sculley2015hidden}.
The \MG serves as a center for Exemplar Machine Learning Models with respect to accuracy, structure, and documentation, to tackle the problem of \ML code \rep. Your \TF implementation should be able to reproduce the results obtained in the original implementation while providing sufficient documentation for companies needing to use the model for industry purposes.

\subsubsection{Internal Review Process}
Before submitting code to the \MG, internally review the repository for functionality and documentation. Our internal review process:
\begin{itemize}
    \item \textbf{Compilation}: Run the model on a CPU, GPU, and TPU to ensure that there are no errors of any kind (complication, execution, etc.) 
    \item \textbf{Training}: Begin the training process on the model with the appropriate training schedule and check if the metrics are behaving correctly.
    \item \textbf{Evaluation}: Load the weights from the original implementation and check if the target metrics are attained.
    \item \textbf{Programming Standards}: Ensure that all code is appropriate formatted to adhere to the \MG's style guide.
    \item \textbf{Reproducibility}: Ensure that all files have proper documentation such as code block explanations, function inputs and outputs, and a detailed README.
    \item \textbf{Consistency}: Ensure that the folder names are identical to the \MG (\cref{garden-table}) and all the files are in their respective folders to ease the merging process during external review.
\end{itemize}

\subsubsection{Repository Documentation}
The objective of having proper documentation is to allow the code to be purposed by many users, and provide them with sufficient instructions and details regarding the functionality of the implementation and its components. 
Gundersen~\etal analyze the relationship between \rep and the documentation requirements and indicate that a well-documented work can largely facilitate \rep~\cite{Gundersen2018ReproducibilityinAI, Gundersen2018ReproducibleAI}.
The main forms of required documentation are: repository README and code comments.

\textbf{Repository README. }
When writing a README for your repository, follow the \href{https://github.com/tensorflow/models/blob/master/.github/README_TEMPLATE.md}{template structure} of the main TFMG repository. In the README, provide instructions for installation, training, and evaluation.
If the model is customizable, \eg allowing the user to choose from several models in the YOLO family, this customization should be discussed in the README.
The README should be reviewed during the final stages of the project, in accordance with the guidelines provided by the \MG.

\textbf{Code Comments. }
Functions should begin with a comment stating the purpose of the function, an explanation of all the inputs required and the outputs returned, along with the shape, rank and type of all the arguments (perhaps given with type annotation).
In addition, each non-trivial code block should be documented with its purpose and the mathematical functions it uses.

\subsection{Tutorials \& Demos}
Before a model is placed in the \TFMG, it should be readily available for public use with the help of demonstrations and tutorials.
Demonstrations show that the exemplar is an accurate model reimplementation.
Tutorials explain how to use the model and what steps to complete before applying it to specific tasks.

\subsubsection{Demonstrations}
When creating a demonstration, all the metrics indicated in the original paper must be presented. To verify that a model is completed, the final demonstration to Google representatives should show that the exemplar implementation's performance matches that of the primary implementation on all the metrics presented in the original research paper.
Our team provided two main demonstrations: a \TF Lite Mobile Application (for Android and iOS) and a video demonstration indicating Frames per Second (FPS) and Mean Average Precision mAP.\footnote{See \url{https://github.com/PurdueCAM2Project/TensorFlowModels/tree/main/yolo/demos/examples}.}

\subsubsection{Creating a Tutorial}
\begin{figure*}[h!]
    \centering
    \includegraphics[width=0.8\textwidth]{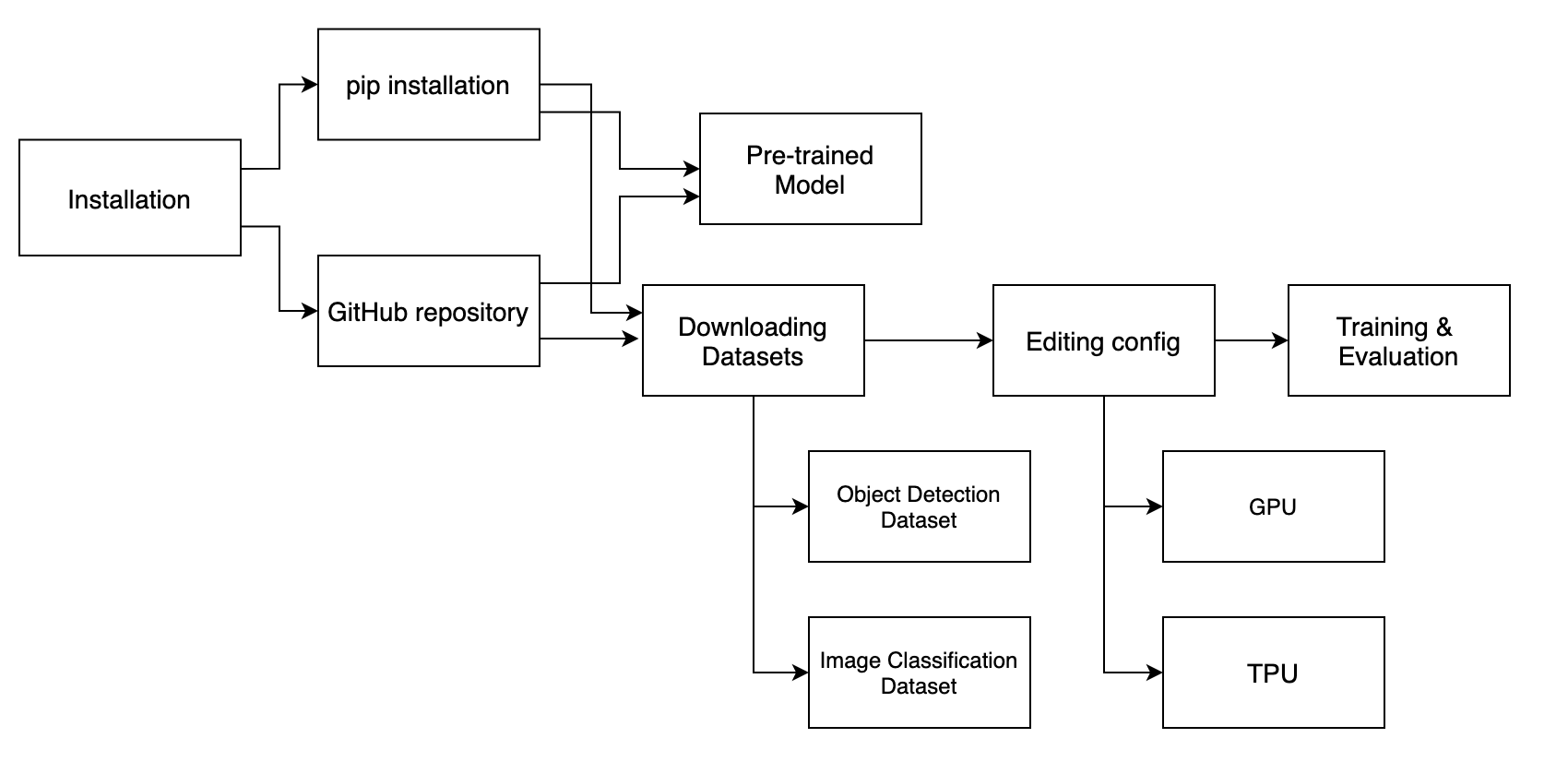}
    \caption{
    \TFDS Tutorial Structure. Each step of this structure must have clear instructions within the tutorial.
    }
    \label{fig:tutorial-struct}
\end{figure*}

Before developing a tutorial, the first step is to create a structure that it must follow. This can done in the form of a flowchart, as shown in \cref{fig:tutorial-struct}. Google has official tutorials on their repository, which can be used as a template when creating a Google Colab or README. Our team has developed a set of two main tutorials to showcase the YOLO model's accurate functioning: \TFDS (TFDS) Usage and YOLO Model Usage Colab tutorial. Throughout the tutorial development process, ensure that all steps are well-documented and simplified, with proper examples and resources.

\textbf{Selecting a demonstration dataset.}
Most tutorials require a dataset to showcase the results. The particular dataset to be used must be carefully selected to ensure the tutorial is accurate from a legality perspective. It is recommended toe select a dataset that is frequently used within the ML community, such as COCO~\cite{COCO} and VOC~\cite{VOC}.

When using a public dataset for the tutorial, licensing must be considered since the tutorial will be available for public access on the \TFMG repository. It must be checked in the License Terms \& Conditions whether or not public distribution of the full dataset or a sub-dataset is allowed for commercial activity. The license must give permission to redistribute and modify the images for free and permanently, such as Creative Commons (CC)~\cite{Wiki:CreativeCommons}. For example: COCO is sometimes difficult to use because its constituent images have varying licenses (only 5\% are unrestricted) and require attribution, with constraints on adaptation and commercial activity usage. For the fields of \CV and \NLP, \TF has several internal datasets (TFDS)\footnote{See \url{https://www.tensorflow.org/datasets}.} that might work just as well. 

In the case of the \TFDS Tutorial, the team did not use ImageNet because the licensing did not allow a subset of this dataset to be created and distributed. ImageNet is frequently used for the metric evaluation of \OD models because of its size, labelling and variety of objects~\cite{krizhevsky2012imagenet}. To replace ImageNet, the OpenImage Dataset (OID) was used, because all images have ``CC BY 2.0'' license, which means they can be shared or adapted and only attribution is required.

\section{Release} \label{release}
\subsection{External Review}
After the model engineering process is completed and internally reviewed by the team, the components must be submitted to Google to publish in the \MG.
This submission is done in phases in the form of GitHub Pull Requests (PR).

\subsubsection{Pull Request Format}
When submitting a GitHub Pull Request to the \TFMG repository, it must provide all the details mentioned in the \href{https://github.com/tensorflow/models/blob/master/.github/PULL_REQUEST_TEMPLATE.md}{provided format}. Each component of the model (Data Pipeline, Model, Loss Function) should be submitted separately to reduce the difficulty of reviewing each PR.
The size of the PR will vary based on the component in the PR.
We suggest submittign two PRs for each model component, one for utility operations and another for the actual component.

\subsubsection{Placement in TFMG}
During the initially accepted PRs, the model will be placed in the appropriate \code{beta} folder within the repository, indicating that it has not been completed.
Upon final approval, the model will be placed in the \code{official} folder within the \TFMG repository, where it can be used and extended by the community. 

\subsection{Final Training \& Publish Weights}
Once the external review is completed and the final model has been published to the \TFMG, it is the team's responsibility to begin the final training process to obtain the weights for the Machine Learning model. After the final weights have been obtained, they must be published along with a pre-trained model (\TF SavedModel format) to the \href{https://tfhub.dev}{\TF Hub}. For Image Classification models, it is required to provide the option of loading the TFHub model with or without the head. Upon publishing the model to the TFHub, it must be named appropriately in accordance with the current labeling system --- the suffix \code{\_vX} where X is the version of the model.
\section{Effective Engineering Practices} \label{effective-eng}

We have begun to identify effective engineering practices for the model exemplar engineering process.
These practices fall under three categories: software reuse, engineering tools, and project management.

\subsection{Software Reuse} \label{appendix-b}
The \TFMG has a modular structure, supporting component re-use between exemplar implementations. Modularity both simplifies implementation, and accelerates innovation: model components can be recombined into a new model performing a different function. For example, the YOLO family is targeted towards object detection, but can be used for image classification by connecting an image classification head to the current backbone. 

\subsubsection{Reusing Code}
Components can be re-used from the basic building blocks of the \TF API, and from higher-level components from other exemplars.
The \TF API~\cite{TFAPIdoc} documents the symbols and operations available within \TF. If your desired operation is not found, look through the \code{ops} folder of other projects within the \MG. When a custom operation has been tested and verified for accuracy, it will be added to a list of all existing \MG custom operations. Once a custom operation is developed by a team and published in the \MG, it can be used by another team within the Gardener community. All custom operations within the \MG have proper documentation and thus, reading through the file comments will help determine if the operation satisfies your requirements or if modification is required. If the file comments are insufficient, then searching through the file for keywords may suffice.

\subsubsection{Writing Reusable Code}
When custom \TF implementation is required, always check if there exists an accurate implementation within the official \TF API or the \TFMG. If the operation are not found, then your team is responsible for constructing the custom operation. It must then be included within the \code{ops} folder, so other developers can easily find your operation if they are looking for something similar. In your custom operations, prefer \TF APIs and Model Garden operations whenever possible to reduce external dependencies. This will ensure that your code can be used by other TFMG developers without imports from outside the Model Garden.

Our team has developed many custom operations, which can be used by any team contributing to the Vision API. As the number of \MG contributors increases, the availability body of reusable code will reduce the need for new code.

\subsubsection{Case Study of Code Reusability}
We have also started work on an implementation of CenterNet in \TFMG based on the Object Detection API (ODAPI) implementation. This experience was different from the YOLO exemplar: there is already a TensorFlow-based implementation from ODAPI, and large pieces of the code can be reused directly without modification.

The challenge is to categorize functions and components of the model based on their reusability in the new API. For example, some functions only require TensorFlow or its dependencies, but others require functions from ODAPI, which cannot be used directly within the \MG without introducing undesirable dependencies. These functions need to be rewritten or adjusted in order to be used in the \MG.

When replicating a neural network, the debugging efforts will depend on the target speed of the model. For example, if the current accuracy is 2-3\% lower than the final target speed of the model, then code optimization will usually not be required, but debugging will end up bridging the performance gap. However, the model purpose will change the debugging approach.

\subsection{Engineering Tools} \label{Tools}
The Exemplar Engineering Problem is difficult, and tools are helpful.
Our team is currently developing software engineering tools for this purpose.


\subsubsection{Linting}
A main aspect of code review is documentation and formatting. Google has specific guidelines for writing code, so our team is currently developing a custom linter using the current \TFMG Pylint config file. This linter will also detect if an operation can be used on a \TPU (TPU) or not, which is described in \cref{portability}.

\subsubsection{VM Monitoring}
For training and testing purposes, having a \VM (VM) can help with computationally expensive tasks. Google Cloud Platform (GCP) is frequently used by ML engineers for model training. Our team is currently working on an automated tool to monitor our GCP account and ensure that the budget is not exceeded. 
This VM monitoring tool will turn the VM on and off based on when a task is started and completed. This tool will also provide warning notifications when the budget has nearly been reached and will automatically shut down the VM when it meets the budget limit.


\subsection{Project Management} \label{PM}
Due to the level of expertise required for publishing a model in the \TFMG, teams are expected to be large.\footnote{We acknowledge that ``man-months'' are mythical. Since our team is composed largely of undergraduate researchers, a larger team permits progress amidst the havoc of the academic semester.} To manage teams of such sizes, proper project management techniques must be practised to minimize idle time and maximize efficiency. The techniques and advice provided below are meant for other institutions thinking of creating an engineering team to join the SIG.

\subsubsection{Team Logistics}
Our engineering team was established in May 2020 and is currently in the External Review stage for the YOLO model family and the Model Engineering stage for the CenterNet Model~\cite{duan2019centernet}. Our team currently has 27 members from \PU (24 undergraduate students, 1 graduate student, 2 faculty advisors) and 2 members from \LUC (1 undergraduate student, 1 faculty advisor). Based on our experience, we recommend a team size of 15 to 25, with a maximum of a 2:1 ratio of new team members to experienced members. Based on our experience, we suggest sub-team sizes of:

\begin{itemize}
    \item Data Pipeline -- 2 members 
    \item Model -- 4 members
    \item Loss Function -- 3 members
\end{itemize}
Our team also calculated the amount of time expected to complete a model with a fixed number of members based on the rate of task completion.
Assuming 15 person-hours per week, the following times are expected to complete a machine learning model:

\begin{itemize}
    \item All Experienced Members: 2.5-3 months
    \item 1:1 New to Experienced Member Ratio: 3.5-5 months
    \item 2:1 New to Experienced Member Ratio: 5-6.5 months
\end{itemize}

\subsubsection{Project Management Tools and Practices}
To keep the team organized and accountable, our team used Monday.com\footnote{See \url{https://www.monday.com}.} for assigning tasks and establishing deadlines. All members are required to document their progress and share their results during a weekly meeting. The leaders of the team should establish short-term and long-term goals.
For example, our short term goal is to place the YOLO model family in the \TFMG, while our long-term goal is to provide a deeper understanding of large scale \ML projects and document best practices for model engineering. 


\subsubsection{Member Onboarding}
When creating an engineering team, members should be selected on the basis of character, technical expertise, and interest.
Successful members have an understanding of software engineering practices, experience in Python, and some training in the principles of machine learning. 
\TF experience will help, but can be learned along the way.

Members will have varying expertise, so a standardized onboarding process should be established to ensure that members are ready for the model engineering process as soon as possible. Our team conducted interviews which tested candidates on their ability to code in Python and apply their existing Python knowledge with \TF to solve a basic \ML problem.

After the interview process, all selected members were given an Onboarding Assignment designed by the experienced members of the team.
The objective of the onboarding assignment is to expose all members to the core aspects of the model engineering process: the Data Pipeline, the Model and the Loss Function.
This assignment used Python software and \TF to solve problems in each of these areas.
Members were then placed on a sub-team based on which part of their assignment worked most efficiently.
The onboarding process will evolve with the team size and structure, but our process can be used as a template during initial stages.

\section{Conclusion}
We hope our experiences help other contributors prepare submissions to the TensorFlow Model Garden.
Through a community effort, we hope to promote machine learning cyberinfrastructure that facilitates open science and the more rapid practical adoption of state-of-the-art techniques.

\section*{Acknowledgments}
We thank Jaeyoun Kim and Abdullah Rashwan for their support and technical expertise. This work was supported by a gift from Google.

\bibliographystyle{ieeetr}
\bibliography{refs/references, refs/JamieMendeley, refs/refs}


\pagebreak
\appendix

\section{Model Components} \label{appendix-a}

Prior to the core ML engineering process for deep learning-based models, all teams must have a basic understanding of neural networks and their operationalization in typical deep learning frameworks.
These components are depicted as shown in \cref{fig:model_structure}.
In these frameworks, a neural network's (model's) architecture is represented as a directed graph for data processing, decomposed into blocks, then layers, and then operations.
We introduce the architecture starting from the most basic to the most complex units.

\footnotetext{For basic \TF resources, refer to the \TF guide at \url{https://www.tensorflow.org/guide}.}

\subsection{Operations}
The \emph{operation} is the basic unit in a neural network. It can be defined as a structure or function; it operates on data, but holds no weights itself. 

\textbf{Differentiability.}
In \TF, these operations can be broken down into 2 fundamental categories: differentiable and non-differentiable. Differentiable operations are those that can be used inside the neural network directly. These are operations that are used to forward propagate or to optimize a neural network. The best way to construct one of these operations is to use the $\code{tensorflow.math}$ API. The majority of these functions have a predefined and optimized inverse or differentiation method that guarantees that any operation that uses them will be differentiable by TensorFlow. Non-differentiable operations can be used in post-processing and pre-processing.

For Differentiable Operations, the best way to test these functions is to use the gradient tape to watch an input tensor and compute the gradient of that tensor. If the function is differentiable, the gradient output put on the tape will not have any values of $\code{None}$ contained within it. 

\textbf{Execution Mode.}
All operations must be usable in both eager and compiled executions. The $\code{tensorflow.math}$ API also ensures that operations will be usable in both eager and compiled execution. Eager execution tells TensorFlow to compute all values at runtime making all operation run slower as each dependency is not know ahead of time. Graph or Compiled Execution is a system that tells the TensorFlow engine to precompute and link all dependencies ahead of time relative to a some place holder input. This essentially stores how to get from input to output for a string of operations, allowing TensorFlow to reference this binary when ever it is needed in order to operate much faster. For more information on how this works and why it is used as on optimization, look into code compilation vs code interpretation vs just in time code compilation.

These operations do not need to be limited to the $\code{tensorflow.math}$ API, but they must still be usable in both graph and compiled executions. If you limit yourself to only TensorFlow operations, that is, any function found in the TensorFlow API, or any built-in Python functions, this should not be an issue. However, if there is ever an instance that requires the use of non-TensorFlow functions, like $\code{numpy}$, the best way to allow graph execution is to use $\code{tensorflow.py\_function}$. Using $\code{tensorflow.py\_function}$ will tell TensorFlow that the operation contained within the function cannot be compiled in graph execution and therefore to switch to eager execution when operating in that code block. This will also give you access to the TensorFlow $\code{numpy}$ conversion methods in the Tensor class of TensorFlow, as these methods are only available to engineers in eager execution.

For all operations in general, the best way to test them for usage in both eager and graph execution, is to wrap the function with the $\code{@tf.function}$ decorator. This decorator will tell the TensorFlow Engine to compile to the subsequent function, allowing you to test the function for graph execution.

\subsection{Layers}

A \emph{layer} is an extension of an operation that is parameterized in some way by a consistent input to a given operation. This layer is the fundamental trainable unit from which all building blocks and models are built from. Typically, this parameterization is the instantiation and storing of weights. In most ML frameworks, layers are used to contain a set of operations that learn about data in a novel way. Typically this is done by assigning weights to each operation. 

As an example of the importance of this parameterization, consider the difference between a convolution operation and a convolution layer.
The convolution \emph{operation} is a mathematical operation that performs a weighted summation using the two input tensors.
The convolution \emph{layer} sets one of the inputs to the convolution operation to be a learnable parameter.
This parameter is called the kernel of the convolution and acts like a filter on the input to the layer.
During evaluation and forward propagation, this parameter is held constant.
During backpropagation, the gradient of the loss function is evaluated with respect to all of the parameters in the model, so that the parameter can be optimized.
In this way, the parameters act differently from inputs, because they can only be updated by the optimizer instead of fed into the model; and act differently from constants, because they are changed during training.

The number of inputs and existence of parameters can make the layer semantically different than the operation, but this is not always the case. Layers can be memoryless and merely wrap the underlying operation. The semantic meaning of the layer within the model determines whether or not to parameterize it.

Every differentiable operation can be encapsulated in a layer through adding parameters as with the convolution example. As mentioned before, non-differentiable operations are not used in the model itself and are used in preprocessing and postprocessing. In addition, multiple operations can be combined together with or without parameters to make a layer. The ultimate distinction between layers and groups of operations is that layers are used in the final model and may have parameters while individual operations are usually used elsewhere (preprocessing and postprocessing).

\subsection{Blocks}

A \emph{block} is an organized set of layers that are common repeating units across many models. The block is used to decompose a model in the same way that functions decompose a program into smaller pieces to eliminate redundancy. This reduces the likelihood that bugs can be introduced in the code and allows for easier debugging of the model.

Blocks are usually relatively small, only consisting of 2-6 layers. Blocks should usually be reusable across many different models because the repeating structure has a unified purpose in the models that doesn't change from model to model. In order to do this, blocks, like layers, still have many parameters in order for the blocks to be relatively flexible and usable across many models.

All blocks required for a particular re-implementation can be defined in one file, and when the block is required within the main model, it will be a simple function call. 

\end{document}